\documentclass[prx,aps,10pt,twocolumn,amsmath,amssymb,superscriptaddress]{revtex4-1}
\usepackage{epsfig,graphicx,graphics,amsmath,amssymb,float}
\usepackage[T1]{fontenc}
\usepackage[latin9]{inputenc}
\usepackage{ dsfont }
\usepackage{amsmath}
\usepackage{amssymb}
\usepackage{xcolor}
\usepackage{amscd}
\usepackage{bm}
\usepackage{psfrag}
\usepackage{bbm} 
\usepackage{color}
\usepackage[bookmarks=true,colorlinks,linkcolor=blue,urlcolor=blue,citecolor=blue]{hyperref}
\usepackage{lipsum,physics}

\usepackage{ulem}
\usepackage{subfigure}

\newcommand{\new}[1]{\textcolor{black}{#1}}

\newcommand{\be}{\begin{equation}}
\newcommand{\ee}{\end{equation}}
\newcommand{\bea}{\begin{eqnarray}}
\newcommand{\eea}{\end{eqnarray}}

\newcommand{\mb}{\mathbf}

\begin{document}

\title{Predicting topological invariants and unconventional superconducting pairing from density of states and machine learning}

\author{Fl\'avio Noronha}
\affiliation{International Institute of Physics, Universidade Federal do Rio Grande do Norte, 59078-970 Natal-RN, Brazil}
\affiliation{Departamento de F\'isica Te\'orica e Experimental, Universidade Federal do Rio Grande do Norte, 59078-970 Natal-RN, Brazil}
\author{Askery Canabarro}
\affiliation{Grupo de F\'isica da Mat\'eria Condensada, N\'ucleo de Ci\^encias Exatas - NCEx,
Campus Arapiraca, Universidade Federal de Alagoas, 57309-005 Arapiraca-AL, Brazil}
\affiliation{Quantum Research Center, Technology Innovation Institute, Abu Dhabi, UAE}
\affiliation{Department of Physics, Harvard University, Harvard University, Cambridge, Massachusetts 02138, USA}
\author{Rafael Chaves}
\affiliation{International Institute of Physics, Universidade Federal do Rio Grande do Norte, 59078-970 Natal-RN, Brazil}
\affiliation{School of Science and Technology, Federal University of Rio Grande do Norte, 59078-970 Natal, Brazil}
\author{Rodrigo G. Pereira}
\affiliation{International Institute of Physics, Universidade Federal do Rio Grande do Norte, 59078-970 Natal-RN, Brazil}
\affiliation{Departamento de F\'isica Te\'orica e Experimental, Universidade Federal do Rio Grande do Norte, 59078-970 Natal-RN, Brazil}

\date{\today}

\begin{abstract}
Competition between magnetism and superconductivity can lead to unconventional and topological superconductivity. However, the experimental confirmation of the presence of Majorana edge states and unconventional pairing currently poses a major challenge. Here we consider a two-dimensional lattice model for a superconductor with spin-orbit coupling and exchange coupling to randomly distributed magnetic impurities.  Depending on parameters of the model, this system may display topologically trivial or nontrivial edge states. We map out the phase diagram by computing the Bott index, a topological invariant defined in real space. We then use machine learning (ML) algorithms to predict the Bott index from the local density of states (LDOS)  at zero energy, obtaining high-accuracy results. We also train ML models to predict the amplitude of odd-frequency pairing in the anomalous Green's function at zero energy. Once the ML models are trained using the LDOS, which is experimentally accessible via scanning tunneling spectroscopy, our method could be applied to predict the number of Majorana edge states and to estimate the magnitude of odd-frequency pairing in real materials.  
\end{abstract}

\maketitle

\section{Introduction}

Topological superconductors have attracted significant attention since the proposal that they can harbor Majorana fermions \cite{Read2000,Kitaev2001}. Several artificial platforms have been put forward and are being investigated with the goal of developing topological qubits \cite{Sato2017,Flensberg2021}. These platforms usually involve hybrid structures that combine magnetism, superconductivity, and spin-orbit coupling.  A prominent example consists of magnetic adatoms deposited on the surface of a conventional superconductor \cite{NadjPerge2013,Braunecker2013,Pientka2013}. In these systems, the ferromagnetically ordered atoms induce in-gap states \cite{Yu1965,Shiba1968,Rusinov1969} which overlap forming a topological $p$-wave superconductor. This setup has been  implemented   in atomic chains \cite{NadjPerge2014,Feldman2017,Schneider2021} as well as in  two-dimensional  (2D) islands  and Shiba lattices \cite{Jian2016,Menard2017,Menard2019,PalacioMorales2019,Soldini2023}. In 2D systems, scanning tunneling spectroscopy (STS) experiments have found evidence for zero-energy states at the edge of the magnetic island, which can be interpreted in terms of propagating  Majorana edge states \cite{Menard2017}. However, in practice it can be challenging to distinguish between edge states of trivial and topological phases, or even between two topological phases with a different number of chiral edge states  \cite{Berthold}.

In parallel to these developments, magnetic impurities in superconductors have been shown to generate odd-frequency pairing  \cite{Kuzmanovski,Perrin,NoronhaDMS,Suzuki2022}. The study of odd-frequency superconductivity started when  Berezinskii  \cite{Berezinskii1974} pointed out that the symmetry of the superconducting state allows for a two-fermion pairing function which is odd under time reversal but even under the product of spatial inversion and spin and orbital permutations.  Since then, states exhibiting this type of unconventional pairing have been identified in many different systems; for reviews,  see Refs. \cite{Tanakareview2012,Linder2019}. For a single magnetic impurity or dilute magnetic impurities in superconductors,  the local density of states (LDOS)  measured by STS can be used to estimate the imaginary part of the odd-frequency local pairing function \cite{Perrin,NoronhaDMS}. Remarkably, odd-frequency pairing has also been linked to the presence of   Majorana bound states  \cite{Daino2012,Asano2013,Kashuba2017,Tsintzis2019,Tanaka2024}. In particular,  the anomalous Green's function near vortices in chiral $p$-wave superconductors shows the same spatial structure as the  LDOS associated with subgap  Majorana modes  \cite{Daino2012}.  
 
In this work, we study the relation between topological invariants, edge states, and odd-frequency pairing in a model for a 2D superconductor with randomly distributed magnetic impurities. Besides being naturally present in superconducting monolayers \cite{Menard2019}, the disorder can be introduced by substitution of magnetic adatoms and lattice reconstruction during the sample growth  \cite{Goedecke2022} or by direct atomic manipulation  \cite{Kuster2022}. Sufficiently strong disorder can drive transitions between topological and trivial phases \cite{Qin,Mascot}. In Ref. \cite{Mascot}, these transitions were analyzed using the Bott index. This topological invariant can be calculated in real space and is equivalent to the Chern number in the homogeneous case  \cite{Hastings,Loring_2010,HASTINGS20111699,Toniolo}. The Bott index has been used to characterize topological phases in several disordered and amorphous systems \cite{Bandres,Agarwala,Huang,Focassio2021,Ghadimi,Dantas}. 

We are interested in how the topological classification based on the Bott index correlates with the LDOS  measured in STS experiments.  To address this question, we use machine learning (ML)   \cite{Carrasquilla2020},  a provably efficient method to classify phases of matter \cite{Huang2022,PRB_100.045129,ChePRB2024,Lewis2024,Rouze2024}. In fact, ML techniques have been successfully applied to identify topological phase transitions \cite{Nieuwenburg2017,Deng2017,Zhang2018,Rem2019,Nieva2019,Scheurer2020,Lustig2020,Claussen2020,Holanda2020,Che2020,Molignini2021,Margalit2022}, to estimate model parameters from numerical or experimental data  \cite{Wang2017,Gentile2021,Koch2023}, and to analyze images generated by scanning probe techniques in quantum materials \cite{Burzawa2019,Iwasawa2022,Matthies2022,Basak2023,Sobral2023}. Here we train ML models to predict the Bott index from the LDOS calculated numerically for our model. We show that a training set restricted to data for a clean superconductor can correctly predict the topological phase diagram for a disordered superconductor. The accuracy of the ML prediction improves if the training set also includes disordered systems. In addition, we show that the imaginary part of the odd-frequency local pairing function shows the same spatial dependence as the LDOS, including signatures of edge states, and a simple linear regression model can predict the former from the latter.

\begin{figure}[t]
        \includegraphics[width=0.9\columnwidth]{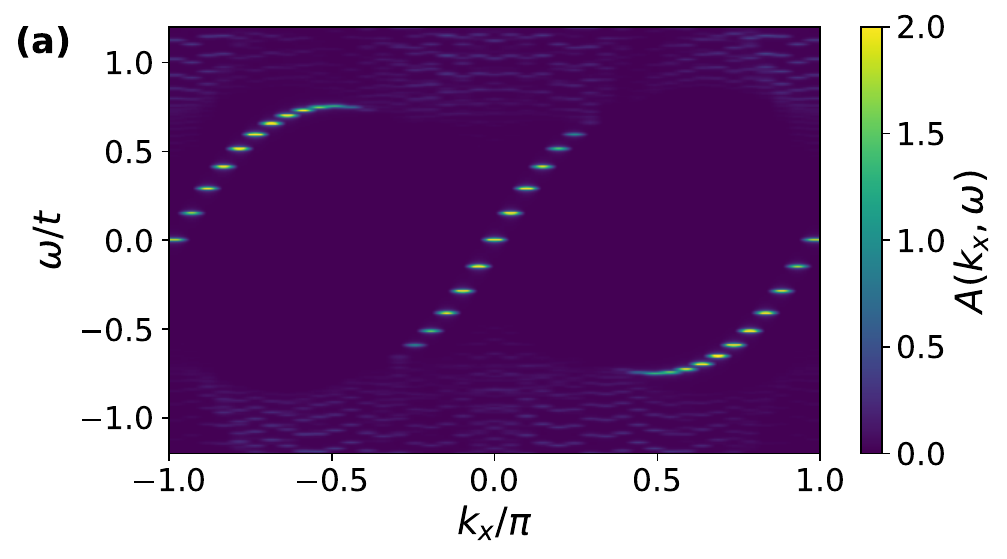}
        \includegraphics[width=0.9\columnwidth]{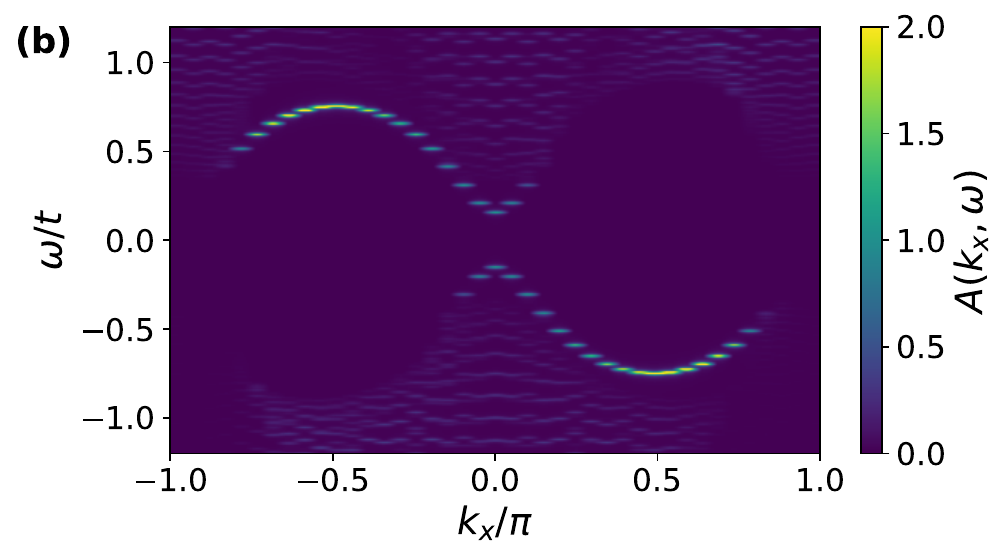}
    \caption{   Dispersion of edge states. We show the edge spectral function $A(k_x,\omega)$ at the lower edge of a clean system ($p=0$) with PBC in the $x$ direction and OBC in the $y$ direction. Here we set $\alpha=0.8t$, $\Delta=1.2t$, $J=2t$, $l_x=l_y=40$, and use the analytical continuation $i\omega\to \omega +i\eta$ with broadening $\eta=0.005t$. (a) For $\mu=-t$, the system is in the topological phase with Chern number $C=2$. The edge spectral function displays two chiral edge modes inside the bulk gap that cross zero energy at $k_x=0$ and $k_x=\pi$. (b) For $\mu=-2t$, the system is the trivial phase, $C=0$, and the edge spectrum is fully gapped. The dispersive modes with the highest magnitude are associated with trivial edge states. } \label{F1} 
\end{figure}

This paper is organized as follows. In Sec. \ref{secmodel}, we introduce the model and discuss trivial and topological edge states in the clean case. In Sec. \ref{secBott}, we describe the phase diagram of the disordered system based on the calculation of the Bott index. In Sec. \ref{secBfromLDOS}, we present our results for the ML prediction of the Bott index from the LDOS. Section \ref{secodd} contains our results on the relation between odd-frequency pairing and the LDOS. We offer some concluding remarks in Sec. \ref{conclusion}.  Finally, Appendix \ref{apA} contains some discussion about  the weights used in the ML model and their relation to features in the LDOS.

\section{General model and clean case \label{secmodel}}

\begin{figure}[t]
\includegraphics[width=0.463\columnwidth]{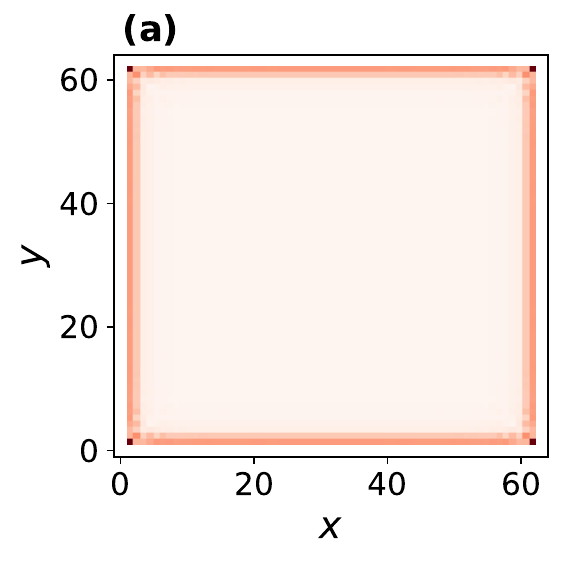}
\includegraphics[width=0.52\columnwidth]{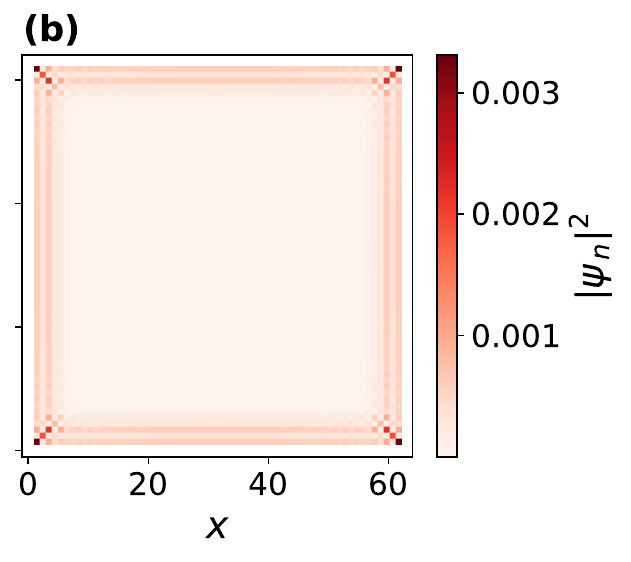}
\caption{  Wave function of in-gap states. We consider a system with OBC and $l_x=l_y=62$. The density plots show the probability $|\psi(x,y)|^2$ in real space associated with the eigenvector of $h$ in Eq. (\ref{Hh}) with the lowest positive energy.  We fix the parameters as in Fig.~\ref{F1}, with two values of the chemical potential: (a) $\mu=-t$ in the topological phase with $C=2$, and (b)  $\mu=-2t$ in the trivial phase with $C=0$. In both cases, the wave function is localized near the edge.  }\label{F2}
\end{figure}

To describe a 2D superconductor coupled to magnetic impurities, we consider the Hamiltonian \cite{Jian2016,Mascot}  \bea
H&=&\sum_{\mb r} \big[ \psi_{\mb r}^\dagger (-t \tau_z+i\alpha\sigma_y\tau_z)\psi_{\mb r+\hat{\mb e}_x}
\nonumber\\
&&+\psi_{\mb r}^\dagger (-t \tau_z-i\alpha\sigma_x\tau_z)\psi_{\mb r+\hat{\mb e}_y} +\text{h.c.}\big] \nonumber\\
&&+ \sum_{\mb r} \psi_{\mb r}^\dagger (-\mu \tau_z+\Delta\tau_x-J_{\mb r}\sigma_z)\psi_{\mb r}.\label{E1}
\eea
Here, $\mb r=x\hat{\mb e}_x+y\hat{\mb e}_y$   denotes positions on a square lattice with primitive vectors $\hat{\mb e}_x$ and $\hat{\mb e}_y$ and lattice spacing set to unity. 
We define the Nambu spinor $\psi_{\mb r}=(c_{\mb r,\uparrow}, c_{\mb r,\downarrow}, c_{\mb r,\downarrow}^\dagger, -c_{\mb r,\uparrow}^\dagger)^{\textrm{T}}$, where $c_{\mb r,\nu}^\dagger$ creates an electron with spin $\nu=\uparrow,\downarrow$ at site $\mb r$. 
The Pauli matrices $\sigma_a$ act in spin space while $\tau_a$ are Pauli matrices in particle-hole space, where $a=x,y,z$. The parameters $t$,  $\mu$, $\Delta$, and $\alpha$ are the hopping, chemical potential, mean-field superconducting pairing, and Rashba spin-orbit coupling, respectively.  We model the magnetic impurities as classical spins ordered in the $z$ direction.  We consider magnetic disorder with a dilution parameter $p$, meaning that every site has a probability $1-p$ to have one magnetic impurity, which interacts locally with conduction electrons via an exchange coupling $J_{\mb r}=J$.  At nonmagnetic sites,  which occur with probability $p$, we set  $J_{\mb r}=0$.

\begin{figure*}[t]
\includegraphics[width=0.65\columnwidth]{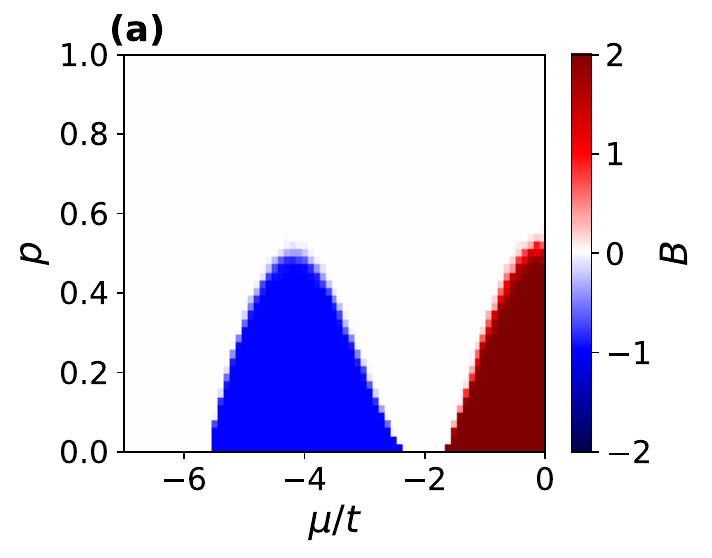}
\includegraphics[width=0.65\columnwidth]{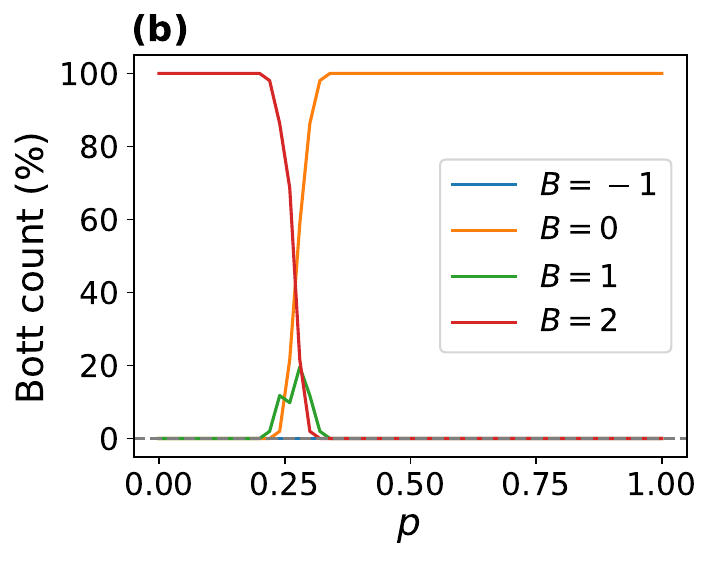}
\includegraphics[width=0.65\columnwidth]{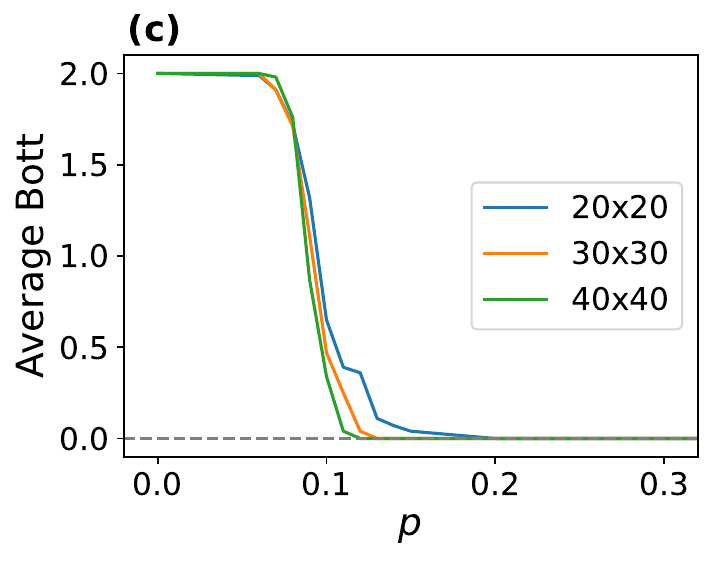}
\caption{   Bott index. 
(a) Topological phase diagram determined by calculating the average Bott index as a function of $\mu$ and $p$.  (b) Percentage of realizations with each value of $B$
for fixed  $\mu=-t$ as a function of $p$.   
In panels (a) and (b) we consider lattices with $l_x=l_y=30$ and $N_d=51$ disorder realizations. 
(c) Bott index for fixed  $\mu=-1.4t$ averaged over $N_d=100$ realizations for different system sizes.  }\label{F3}
\end{figure*}

The Hamiltonian  for a lattice with $N$ sites can be rewritten as \be
H=\Psi^\dagger h \Psi, \label{Hh}
\ee
where $h$ is a $4N\times 4N$ matrix and $\Psi$ is a Nambu spinor with $4N$ components labeled by position, spin and particle-hole index. We diagonalize $h$ for square lattices of linear sizes $l_x$ and $l_y$   in the $x$ and $y$ directions, respectively; hence, $x\in\{0,1,\cdots,l_x-1\}$, $y\in\{0,1,\cdots,l_y-1\}$, and $N=l_xl_y$. We consider both open boundary conditions (OBC) and periodic boundary conditions (PBC). Once we obtain  the complete set of eigenvalues and eigenvectors, we can calculate  quantities of interest, such as  the matrix elements of the electron Green's function \be
\mathbbm G_{mn}(\tau_1,\tau_2)=-\left\langle T_\tau \Psi^{\phantom\dagger}_m(\tau_1)\Psi_n^\dagger(\tau_2)\right\rangle,
\ee
where $m,n\in\{1,\cdots,4N\}$ and $T_\tau$ denotes time ordering with respect to imaginary time. In particular, the normal Green's function contains elements of the form \be
G_{\nu_1\nu_2}(\mb r_1,\mb r_2;\tau_1,\tau_2)=-\left\langle T_\tau c_{\mb r_1,\nu_1}^{\phantom\dagger}(\tau_1)c_{\mb r_2,\nu_2}^\dagger(\tau_2)\right\rangle,
\ee
whereas the anomalous Green's function contains the pairing amplitudes given by
\be
F_{\nu_1\nu_2}(\mb r_1,\mb r_2;\tau_1,\tau_2)=-\left\langle T_\tau c_{\mb r_1,\nu_1}^{\phantom\dagger}(\tau_1)c_{\mb r_2,\nu_2}^{\phantom\dagger}(\tau_2)\right\rangle.\label{anomalous}
\ee

In the homogeneous case where all sites are magnetic ($p=0$), we can diagonalize the Hamiltonian in momentum space and determine the dispersion relation analytically. For $\Delta<J$ and nonzero spin-orbit coupling,  topological   transitions   occur  at critical values of the chemical potential where the gap closes, given by \cite{Jian2016}
\bea
\mu_{ij\lambda}=2(i+j)t+\lambda\sqrt{J^2-\Delta^2},\quad i,j,\lambda=\pm 1.\label{E2}
\eea
In the regime   $2t>\sqrt{J^2-\Delta^2}$, we can reach three phases by varying the chemical potential: a trivial phase with Chern number $C=0$ and two topological phases with  Chern numbers $C=+2,-1$.

Throughout  this paper we set $\hbar=1$ and use the parameters $\alpha=0.8t$, $\Delta=1.2t$ and $J=2t$.
This choice will allow us to compare some of our results with those presented in Ref. \cite{Mascot}.

We start by analyzing the edge modes in a clean system with  $p=0$. \new{Note that here  ``clean'' refers to  the absence of disorder, not  the absence of magnetic impurities.}  For this purpose, we first consider a lattice with PBC in the $x$ direction and OBC in the $y$ direction. Using the translational invariance along   $x$, we define \be
G_{\nu\nu}(k_x,y;\tau)=\sum_{x=0}^{l_x-1}e^{-ik_xx}G_{\nu\nu}(x\hat{\mb e}_x+y\hat{\mb e}_y,y\hat{\mb e}_y;\tau,0), 
\ee 
where $k_x$ corresponds to the momentum in the $x$ direction. We then take the Fourier transform to  Matsubara frequencies and the analytical continuation  $i\omega \to \omega+i\eta $ to obtain the retarded Green's function $G^{\rm ret}_{\nu\nu}(k_x,y;\omega)$.  \new{To analyze the spectrum of the finite-size system, we use small but finite values of  $\eta$,  corresponding  to a phenomenological broadening as used, for instance, in Ref. \cite{Perrin}.} In Fig.~\ref{F1} we show the edge spectral function $A(k_x,\omega)=-\frac1\pi\sum_{\nu=\uparrow,\downarrow}\text{Im}G^{\rm ret}_{\nu\nu}(k_x,y=0;\omega)$ at the lower edge of the system. In Fig. \ref{F1}(a), the chemical potential lies in the topological phase with   $C=2$. As expected, we see two gapless dispersing edge states inside the bulk gap in this case. By contrast, gapless edge modes are absent in Fig. \ref{F1}(b), corresponding to a value of chemical potential  in the trivial phase. 

In Fig. \ref{F2} we consider a lattice with OBC in both directions and plot the probability density $|\psi(x,y)|^2$ corresponding to the wave function of the eigenstate with the lowest positive energy. Remarkably,   both cases with  $C=2$ in Fig. \ref{F2}(a) and $C=0$ in Fig. \ref{F2}(b) indicate the presence of edge states, but the second case is topologically trivial. This example illustrates that even in a clean superconductor it can be challenging to determine whether the system is in a topological or trivial phase based solely on the presence of edge states. In the following, we will turn to the effects of magnetic disorder.

\section{Bott index and disorder-induced topological transitions\label{secBott}}

\begin{figure*}[t]
\includegraphics[width=0.3\textwidth]{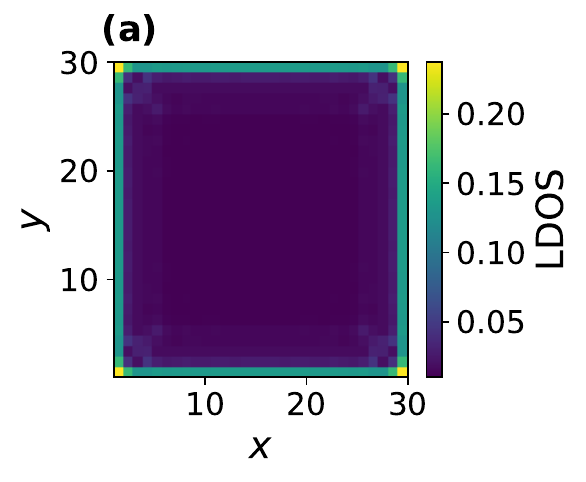}
\includegraphics[width=0.7\columnwidth]{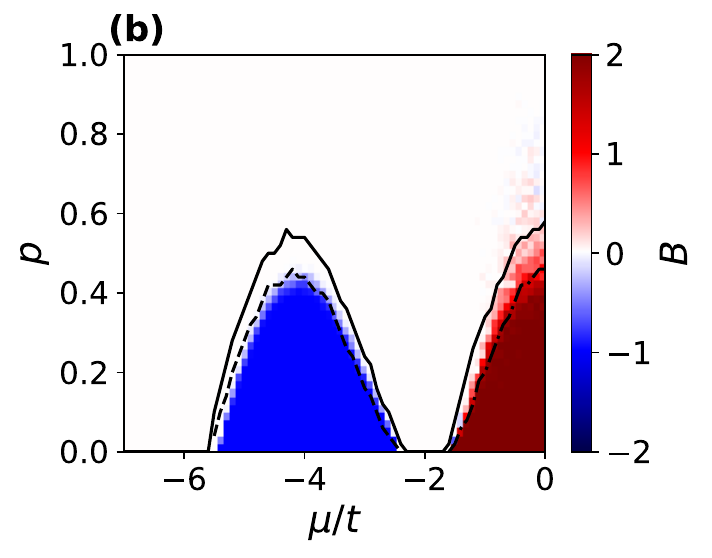}
\includegraphics[width=0.7\columnwidth]{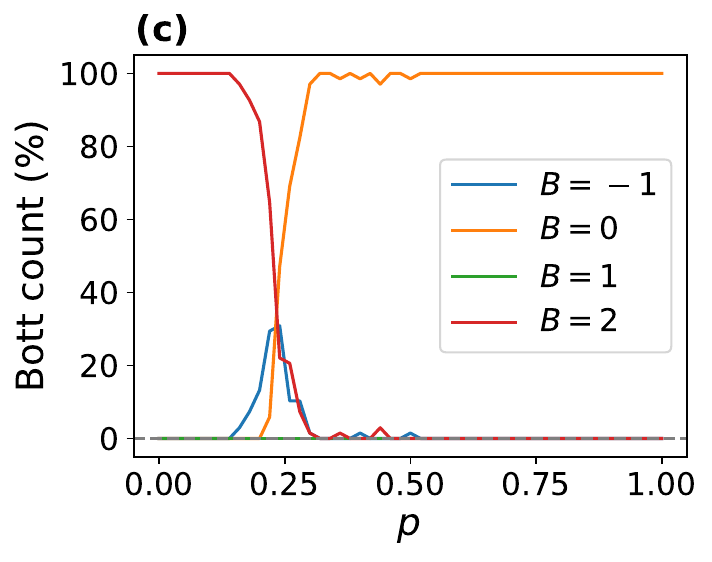}
\caption{ML prediction trained with the LDOS of clean systems. (a) As an example of the training data, we show   $\rho(\mb r,0)$  for the clean system with $\mu=0$, which corresponds to  the topological phase with $B=2$. Here we set the  broadening $\eta=0.04t$.  (b) Phase diagram predicted from the LDOS using  $30\times 30$ lattices with OBC, using a training set with  $p=0$ and $701$ values of $\mu$.  The ML model consists of a neural network with no hidden layer, 3 exit nodes, 100 epochs, the Adam optimizer, sparse categorical  cross-entropy as the loss function, and softmax for the activation function.   The color scale represents the predicted value of  $B$ averaged over  $N_d=68$ realizations.  The solid   lines were extracted from Fig.~\ref{F3}(a) and represent the value of $p(\mu)$ above   which every disorder realization has $B=0$. Below the dashed (dot-dashed) line, every disorder realization has  $B=-1$ ($B=+2$). (c) Counting of the predicted values of the  Bott index as a function of $p$ for fixed $\mu=-t$.  }\label{F4}
\end{figure*}

In the presence of disorder ($p>0$), we use the Bott index to identify topological phase transitions \cite{Hastings,Loring_2010,HASTINGS20111699}. 
To introduce the Bott index, let us consider the eigenvectors of $h$ in Eq. (\ref{Hh}) in ascending order of their eigenvalues. Charge conjugation symmetry implies that for each eigenenergy $\varepsilon$ there exists an eigenenergy $-\varepsilon$.   Let $P$ be the projector onto the subspace spanned by the $2N$  eigenvectors with negative eigenvalues, which correspond to occupied single-particle states in the ground state of the many-body problem.  Using  the $x$-coordinate position operator $X$ and the $y$-coordinate operator $Y$, we define $\tilde{U}=P e^{i2\pi X/l_x} P$ and $\tilde{V}=P e^{i2\pi Y/l_y} P$. We then define $\hat{U}$ ($\hat{V}$) as a $2N\times 2N$ matrix containing the first subblock of $\tilde{U}$ ($\tilde{V}$),  corresponding to the subspace of occupied states. For computational stability, we perform a singular value decomposition $\hat{U}=L_U D_U R_U^\dagger$ and define $U=L_U R_U^\dagger$, where $L_U$ and $R_U$ are unitary matrices and $D_U$ is diagonal. Similarly, $\hat{V}=L_V D_V R_V^\dagger$ and we define $V=L_V R_V^\dagger$. The   Bott index is then defined as \be
B=\frac1{2\pi}\textrm{Im}\{ \textrm{Tr}[ \textrm{Log}(UVU^\dagger V^\dagger) ]   \}.
\ee
This is an integer topological invariant, $B\in\mathbb Z$, equivalent to the Chern number in homogeneous systems \cite{Toniolo}.  We have computed $B$ on a lattice with PBC where all sites are magnetic and found perfect agreement with the phase diagram determined in Ref. \cite{Jian2016}   using the Chern number.

Next,  we compute the Bott index for disordered systems by averaging over many disorder realizations at fixed values of $\mu$ and $p$. The result is shown in Fig.~\ref{F3}(a).  In agreement with   Ref. \cite{Mascot}, we find two topological phases characterized by  $B=-1,+2$ surrounded by a trivial phase  ($B=0$).  This means that the topological phases present in the clean system are robust against some degree of disorder.

In turn, Fig.~\ref{F3}(b) shows how often each value of the  Bott index appears for a fixed $\mu=-t$ as we vary $p$  across the disorder-induced topological transition. For small $p$, all disorder realizations have $B=2$, while for large $p$ all of them have $B=0$. A critical value of $p$ can be estimated as the point where the fractions of disorder realizations with $B=0$ and $B=2$ coincide.  Interestingly, near the transition we observe a significant fraction of realizations with an intermediate Bott index, $B=1$. We have observed this intermediate Bott index  all along the transition between  $B=2$ and  $B=0$ phases in the phase diagram of Fig. \ref{F3}(a). 

To assess finite-size effects, in Fig. \ref{F3}(c) we show the averaged Bott index as a function of $p$ for different lattice sizes. Clearly, the topological transition from $B=2$ to $B=0$ becomes sharper as we increase the system size.  However, within the finite-size limitations of our numerics, we are unable  to determine whether the transition becomes infinitely sharp and the number of realizations with  $B=1$ vanishes in the thermodynamic limit.  
 A possible alternative scenario is the emergence of an intervening gapless phase,  as found in disordered topological insulators \cite{Sbierski2014,Fu2023}.

\begin{figure*}[t!]
\includegraphics[width=0.65\columnwidth]{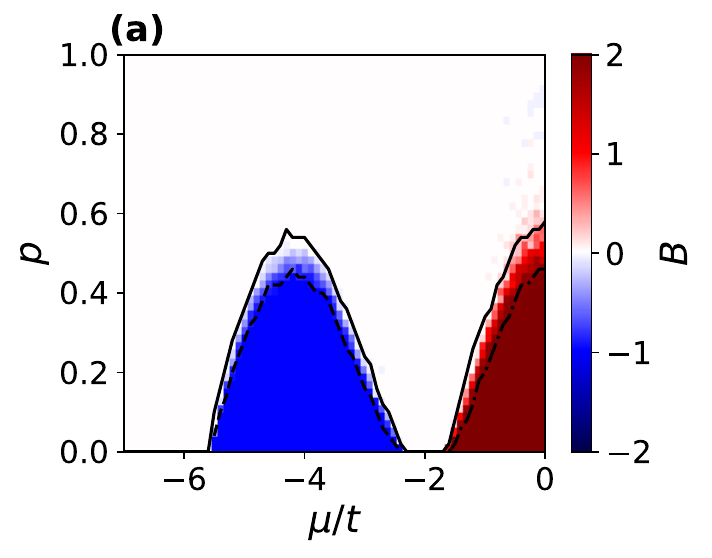}
\includegraphics[width=0.65\columnwidth]{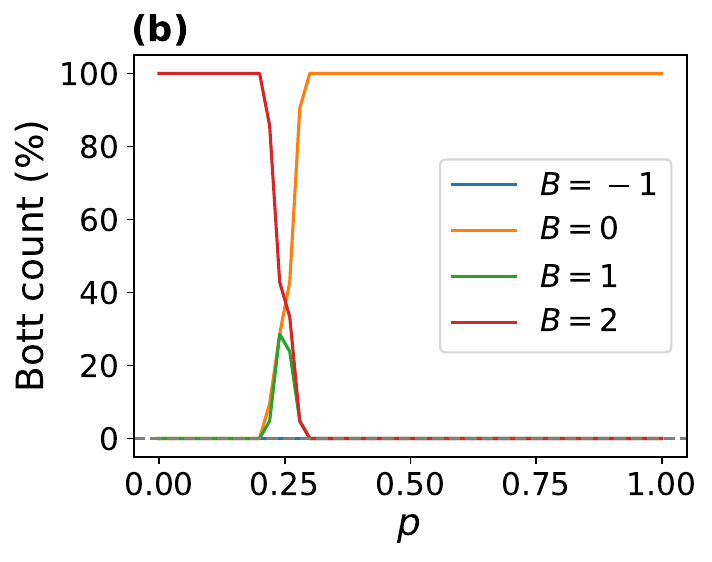} 
\includegraphics[width=0.65\columnwidth]{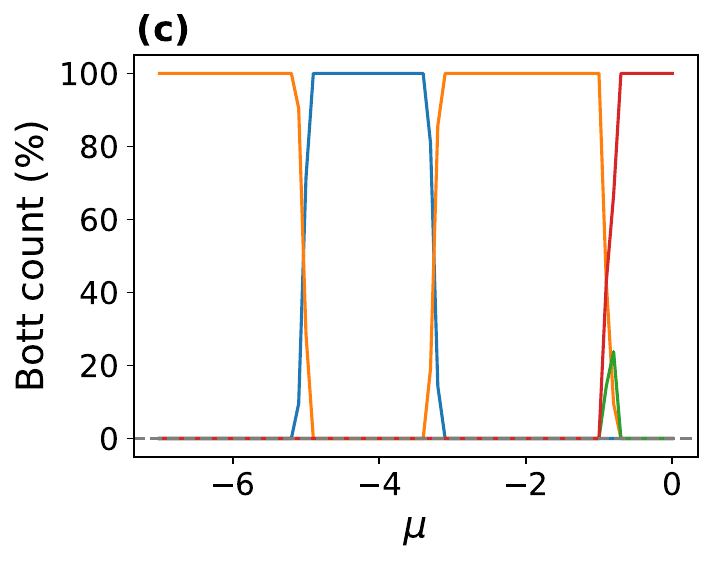} 
\caption{  ML prediction trained with disordered systems. 
 Here we  trained  an ML  model using data for a $71\times 51$ grid in the $\mu\times p$ phase diagram of Fig. \ref{F3}(a). The training set has $N_d=47$ realizations for each   value of $\mu$ and $p$. With this model,  we predict $B$ for a validation set of $N_d=21$ new realizations for each $\mu$ and $p$. Panel (a) shows the predicted $B$ averaged over $N_d$ realizations. The solid  and dashed lines are the same as  in Fig.~\ref{F4}(b).
In panel (b) we fix $\mu=-t$ and count how often each value of $B$ appears. \new{In panel (c), we show multiple phase transitions by fixing  $p=0.3$ and varying $\mu$. }
Here we used a neural network with two hidden layers with $600$ nodes each, rectified linear unit as activation function, and 5 epochs.
To allow any integer as output, we considered an exit layer with a single node, after which we round the result to the nearest integer. We use mean squared error as  loss function. 
When choosing the training set, we excluded data points near the transition, in the region between solid and dashed or dot-dashed lines.
 The accuracy in the validation set is above 99.5\% outside this transition region. }\label{F5}
\end{figure*}

\section{Predicting  the Bott index from the LDOS \label{secBfromLDOS}}

The main spectroscopic signature of  Majorana bound states is a zero-energy signal inside the superconducting gap as measured by STS \cite{Berthold}.  For instance, Majorana bound states at the edges of a one-dimensional  superconducting wire lead to zero-bias peaks in the LDOS  near the edges. In the case of chiral Majorana modes at the edges of a 2D superconductor, their linear dispersion [seen in Fig.~\ref{F1}(a)] gives rise to a flat contribution to the LDOS inside the gap.  However, other mechanisms such as Andreev bound states can also produce similar signatures,  in a way that unambiguous experimental evidence for Majorana bound states is still lacking. In this section, we show how ML can be used to distinguish between Majorana bound states and trivial edge states, such as the one shown in Fig.~\ref{F2}(b).  To generate the training data, we calculate the LDOS\be
\rho(\mb r,\omega)=-\frac1{\pi}\sum_{\nu=\uparrow,\downarrow}\text{Im}\, G^{\rm ret}_{\nu\nu}(\mb r_1=\mb r,\mb r_2=\mb r;\omega),
\ee
for positions $\mb r$ on a lattice with OBC. We use numerical results for the LDOS  at frequency $\omega=0$  to train  ML models to predict the Bott index.

Recall that Eq.~(\ref{E2}) provides us with the critical chemical potentials where topological phase transitions take place in the clean superconductor. Therefore, we know $B$ for each point along the $p=0$ axis of the phase diagram in Fig. \ref{F3}(a). Our first task is to use the LDOS of these clean superconductors with their corresponding $B$ [see Fig. \ref{F4}(a) for an example] to train 
 an ML  model to predict $B$ for disordered superconductors with $p>0$. \new{For this purpose, we use a neural network with no hidden layer and  three exit nodes.   The exit layer is influenced by a softmax activation function, which causes the three exit neurons to provide probabilities that the data belongs to each of the three  classes labeled by the Bott index. In the absence of hidden layers, the neural network attempts to learn a linear function  that maps the LDOS to one of the three classes. By analyzing the weights that the neural network assigns to the LDOS for all lattice sites in different phases (see Appendix \ref{apA}), we notice that in the trivial phase the weights tend to be higher in the bulk, whereas the weights associated to positions along the edge become higher in the topological phases. We interpret these weights as a sign that the ML model learns to identify patterns in the edge states, which are then associated with the different values of the Bott index. }

 In Fig.~\ref{F4}(b), we  show the disorder-averaged prediction for  $B$ as a function of $\mu$ and $p$, \new{based on training the ML model with data along the $p=0$ line.} For comparison, we also show lines extracted from the direct calculation of the Bott index in   Fig.~\ref{F3}(a).   The solid   lines represent the   values of $p$ above   which every disorder realization has $B=0$, whereas below the dashed (dot-dashed) line in the left (right) dome every realization has $B=-1$ ($B=2$).  \new{Although this model was trained only at $p=0$, it can predict qualitatively well the average value of $B$ at finite disorder strength. This result is surprising given that disorder  leads to complex spatial patterns that  differ significantly from the  clean case.} In Fig. \ref{F4}(c),  we fix $\mu=-t$ and show how often each predicted value of the  Bott index appears as a function of $p$. Since the value $B=+1$ does not occur  in the training set for $p=0$,  this classification model is not able to predict the $B=+1$ realizations seen in Fig.~\ref{F3}(b). Moreover,  it incorrectly predicts $B=-1$ realizations near the transition between the topological phase with $B=2$ and the trivial phase. Thus, this method provides reasonable predictions for  the average $B$ but not for the frequency of the possible values of  $B$  near the topological transition.

\begin{figure*}[t]
\includegraphics[width=0.75\textwidth]{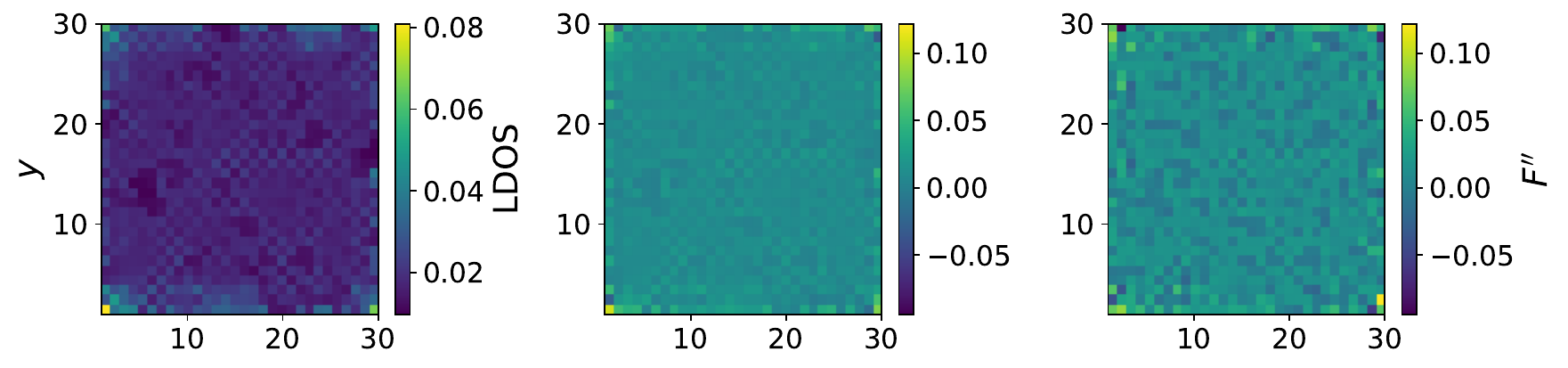}
\includegraphics[width=0.75\textwidth]{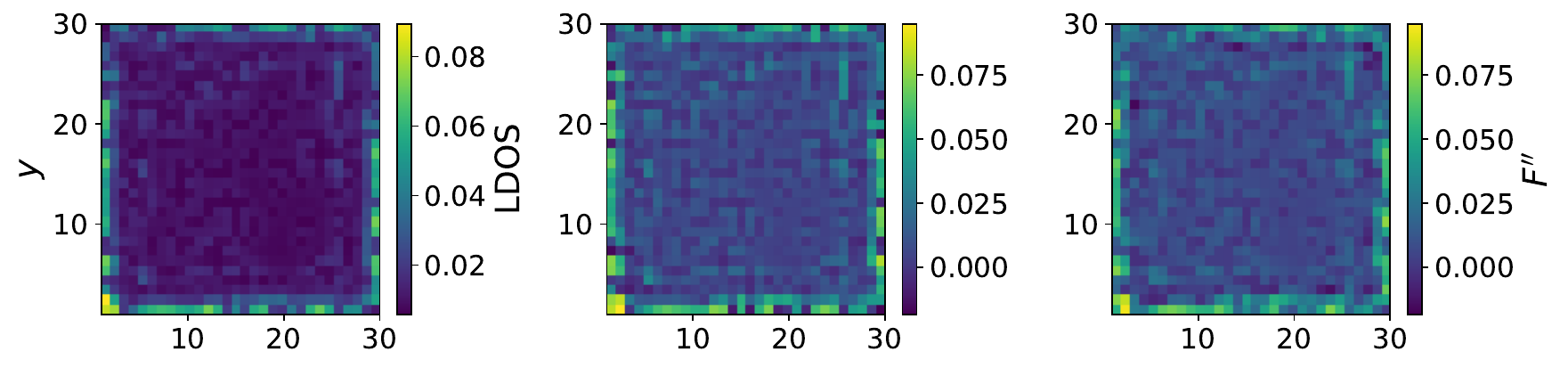}
\includegraphics[width=0.75\textwidth]{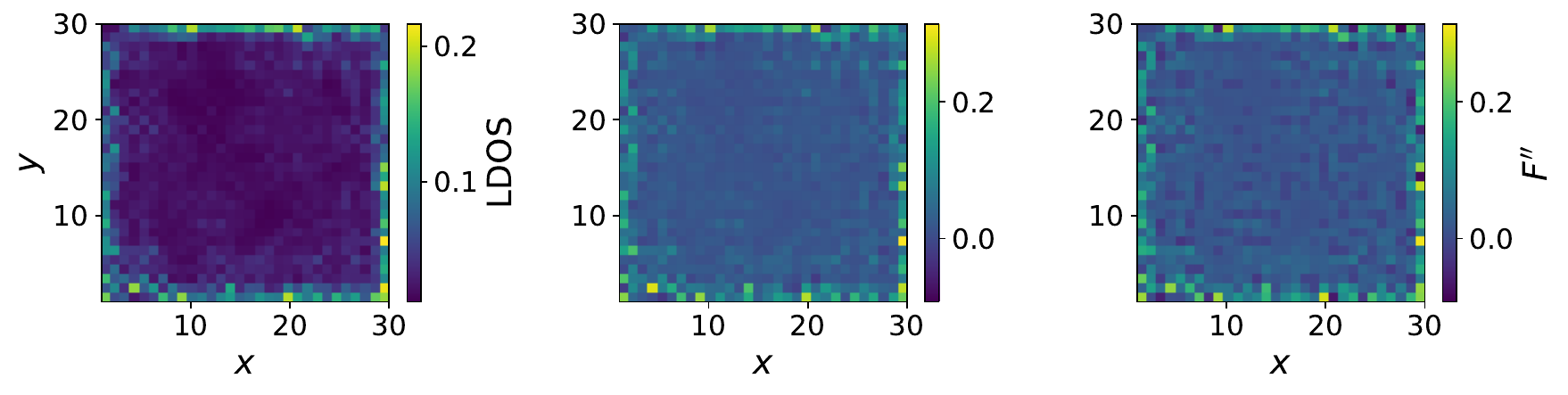}
\caption{ Predicting odd-frequency pairing from the LDOS. Here we train an ML   model with the LDOS at zero frequency, $\rho(\mb r,\omega=0)$, and the corresponding ${F''(\mb r,\omega=0)}$ of $N_d=47$ disorder realizations for several values of $\mu$ and $p$. We then use this model to predict $F''(\mb r,\omega=0)$ from  $\rho(\mb r,\omega=0)$ of a new testing set with $21$ disorder realizations.   The panels on the left show the LDOS, the ones in the middle show the corresponding $F''(\mb r,\omega=0)$ predicted with the ML model, and the ones on the right show the exact value of $F''(\mb r,\omega=0)$.
The   three rows are realizations of $p=0.3$ with $\mu/t=-2,-4,0$, corresponding to  $B=0,-1,+2$, respectively.}\label{F6}
\end{figure*}

\new{Next, to improve the accuracy of the model, we include   realizations with $p>0$ in the training set, but we avoid the region near the transition.  Moreover, we now  consider a neural network with  two hidden layers where the exit neural layer has a single node.  The hidden layers  enable the network to learn complex patterns and relationships between the input and output data. The   single neuron without softmax activation function in the exit layer  allows us to obtain any real number as output, which is then rounded to the nearest integer.  This method  is  able to predict a Bott index not included in the training set $\{-1,0,+2\}$. 
 The idea is to  check whether this model finds the $B=+1$ realizations without the inclusion of  any realization with this topological invariant in the training set.} The result of this approach is shown in Fig.~\ref{F5}. Figure  \ref{F5}(a) shows the average of $B$ over disorder realizations. Clearly,  this result is better than  Fig.~\ref{F4}(b) since the predicted topological transitions now   take  place between the solid and dashed lines extracted from Fig. \ref{F3}(a). The accuracy of this model outside the transition region is higher than $99.5\%$. Figure \ref{F5}(b) shows the Bott indices counting for $\mu=-t$ as a function of $p$. \new{In Fig.  \ref{F5}(c) we can see multiple phase transitions as we vary $\mu$ while fixing the value of $p$.} Remarkably,  this model  predicts $B=+1$ realizations near the topological transition between $B=0$ and $B=+2$, in agreement with the direct calculation of the Bott index. We note, however, that the counting of predicted $B=+1$ here is only qualitatively similar to the  counting shown in Fig.~\ref{F3}(b).

\section{Odd-frequency pairing \label{secodd}}

We now turn to the anomalous Green's function.  Taking the Fourier transform of Eq. (\ref{anomalous}) and the analytical continuation $i\omega\to \omega +i\eta$,  we obtain the retarded anomalous Green's function $F^{\rm ret}_{\nu_1\nu_2}(\mb r_1,\mb r_2;\omega)$. We focus on  local correlations, $\mb r_1=\mb r_2=\mb r$, with  $\nu_1=\downarrow$ and $\nu_2=\uparrow$. To simplify the notation, we define $F(\mb r,\omega)=F^{\rm ret}_{\downarrow\uparrow}\big(\mb r,\mb r;\omega\big)$. Since this correlation is local, it is an even-parity $s$-wave correlation. The pairing function $F(\mb r,\omega)$ may be decomposed into even-frequency  singlet $F^e(\mb r,\omega)$ and odd-frequency triplet $F^o(\mb r,\omega)$ components in the form \cite{Perrin,NoronhaDMS}\be
F^{e/o}(\mb r,\omega)=\frac12[F(\mb r,\omega)\pm F^*(\mb r,-\omega)].
\ee
As a consequence, the real and imaginary parts of $F(\mb r,\omega)$ satisfy $F'(\mb r,0)=F^e(\mb r,0)$ and  $iF''(\mb r,0)=F^o(\mb r,0)$, respectively. This means that, at zero energy,  the real component of the local anomalous Green's function corresponds to even-frequency pairing, but a finite imaginary part reveals the presence of odd-frequency pairing. 

\begin{figure*}[t]
\includegraphics[width=0.65\columnwidth]{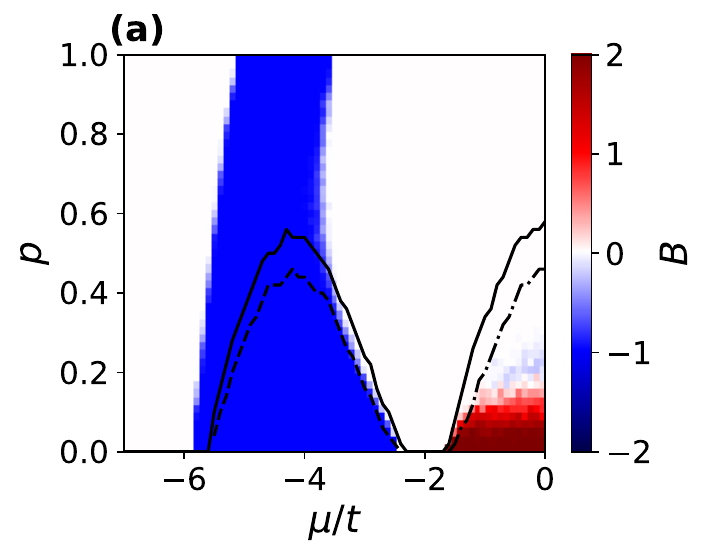}
\includegraphics[width=0.65\columnwidth]{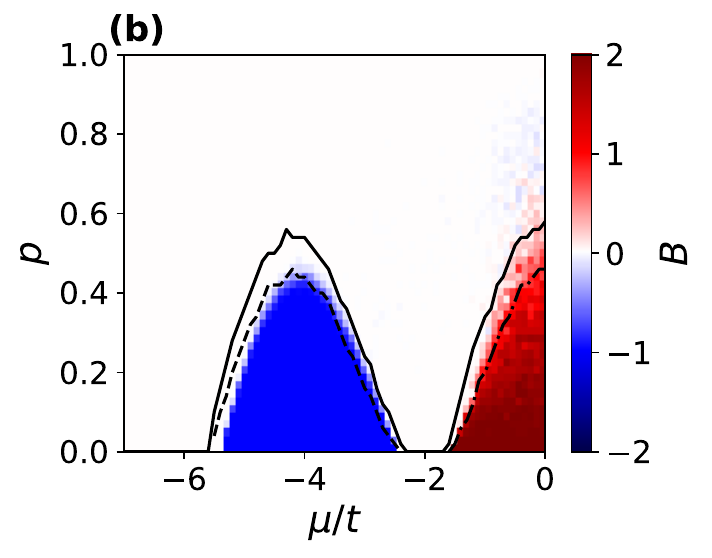}
\includegraphics[width=0.65\columnwidth]{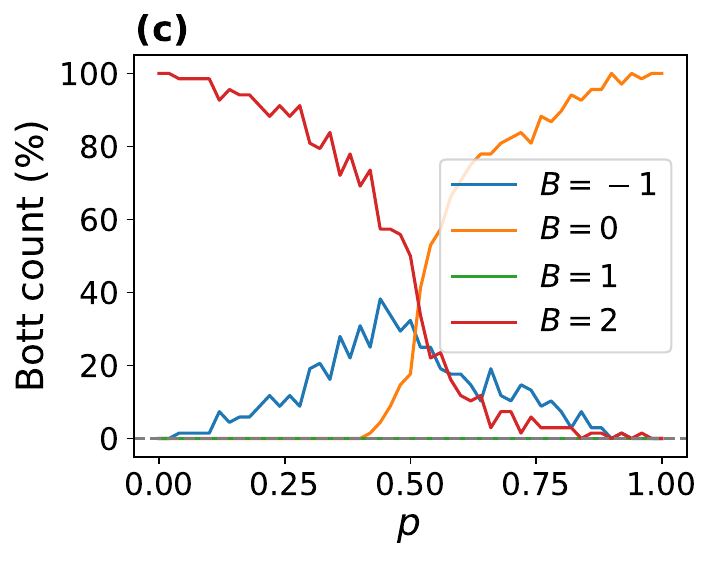}
\caption{  ML prediction trained with the   anomalous Green's function in clean systems. (a) Phase diagram predicted from the real part $F'(\mb r,0)$ in a Random Forest model with 1000 estimators. Different ML methods for $F'(\mb r,0)$ in clean systems led to even worse results. (b) Phase diagram predicted from  the imaginary part $F''(\mb r,0)$ using  a neural network with no hidden layer, 25 epochs and 3 exit nodes. The models were trained using $N_d=701$ values of $\mu$ and predicted $N_d=68$ realizations for each value of $\mu$ and $p$. (c) Counting of the Bott index predicted from $F''(\mb r,0)$ as a function of $p$ for fixed $\mu=-t$. }\label{A1}
\end{figure*}

The model of Eq.~(\ref{E1}) has a superconducting term $\Delta$ that induces a singlet, $s$-wave, even-frequency pairing, which survives in the presence of disorder. However, the interplay between this conventional superconductivity and magnetism leads to the emergence of odd-frequency pairing~\cite{Kuzmanovski}, which we verify by checking that $F''(\mb r,0)$ is finite. 

\subsection{Predicting the anomalous Green's function from the LDOS}

While the pairing function is not directly measured in experiments, it can be estimated  using the LDOS  in the case of a single magnetic impurity or dilute magnetic impurities in superconductors~\cite{Perrin,NoronhaDMS}. Motivated by this fact, here we train a ML model of linear regression using $\rho(\mb r,0)$   to obtain $F''(\mb r,0)$. The training set consists of $N_d=47$ disorder realizations for each value of $\mu$ and $p$. The testing set has $N_d=21$ new realizations for each $\mu$ and $p$, for which we predict $F''(\mb r,0)$ using only the corresponding  LDOS. Comparing the predicted results with the actual functions, we find that  the root-mean-square error of the set of all 21 realizations is $\delta=0.0083$. In Fig.~\ref{F6} we show the LDOS, the  ML prediction and the actual $F''(\mb r,0)$ of a realization with $p=0.3$ and three different values of $\mu$ corresponding to $B=0,-1,2$.   With the simple linear regression method used here,  we conclude that the predicted functions have reasonable agreement with the actual odd-frequency pairings at zero energy. Neural networks with hidden neuron layers may be used in order to improve accuracy. However, since the exit layer  would now have 900 nodes corresponding to each position of the $30\times 30$ lattice, higher computational resources would be necessary.

\subsection{Predicting the Bott index from the anomalous Green's function}

Since the ML models used to predict $B$ from LDOS worked well and there is an interesting relation between LDOS and $F''(\mb r,0)$, we also trained ML models to predict $B$ from the anomalous Green's function. Similarly to the result in Fig.~\ref{F4}, in Fig.~\ref{A1} we start by using clean systems ($p=0$) to train ML models.   With the trained models we use the pairing function of several realizations with disorder  to predict the topological phase diagram.  

In Fig.~\ref{A1}(a) we show the result obtained when we use a Random Forest model trained with the real part of the anomalous Green's functions, $F'(\mb r,\omega=0)$. The result clearly fails because it incorrectly  predicts that the topological phase with  $B=-1$ extends all the way to $p=1$. We tried several other ML models, including neural networks, and all of them predicted  topological phase diagrams in poor agreement with the result in Fig. \ref{F3}(a). This is in contrast with the result discussed in Fig.~\ref{F4}(b), where ML trained with the LDOS of clean systems was enough to predict qualitatively well the phase diagram. \new{To interpret this result, we note that the real and imaginary parts of (normal or anomalous) Green's functions are related by Kramers-Kronig relations. However, these relations involve an integral over the full energy spectrum. For this reason, the real part of the anomalous Greens function $F'(\mb r,\omega=0)$ is influenced by the imaginary part $F''(\mb r, \omega)$ for every energy $\omega$. As a consequence, while $F''(\mb r, \omega=0)$ shows clear signatures of topological edge states, we do not observe these signatures in $F'(\mb r, \omega=0)$. This explains why the prediction with the real part in Fig. ~\ref{A1}(a) is so poor. It also indicates that the main feature that the neural network uses to classify the images is the enhancement of the LDOS near the edges in the topological phases. }

\begin{figure}[t]
\includegraphics[width=0.9\columnwidth]{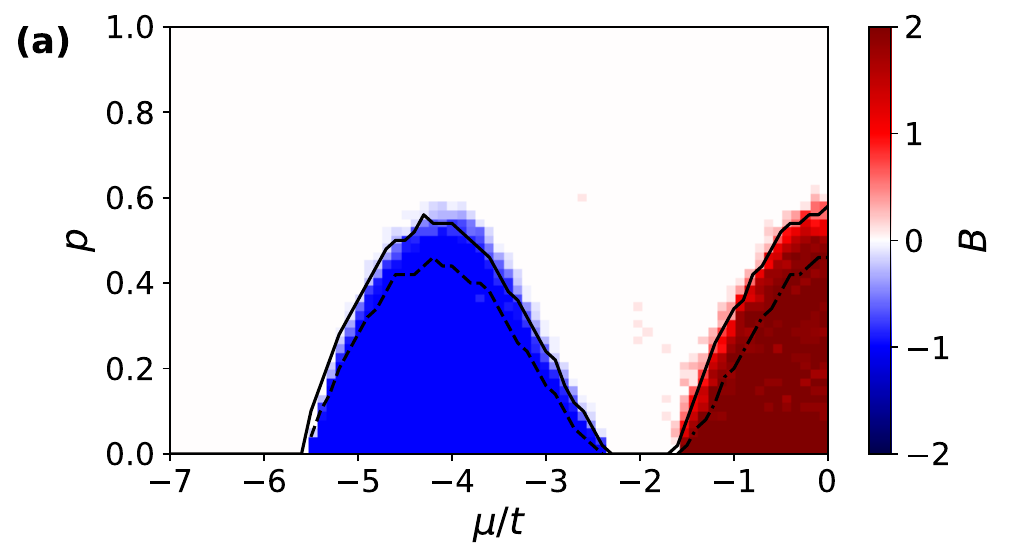}
\includegraphics[width=0.9\columnwidth]{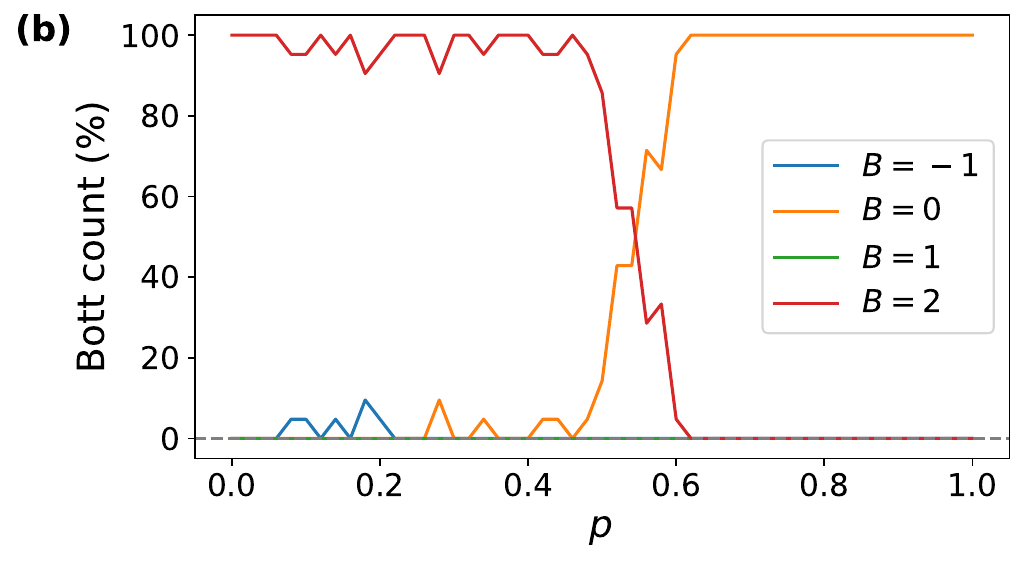}
\caption{ML prediction trained with the real part of the anomalous Green's function in disordered systems. (a) Phase diagram for the average Bott index  predicted from $F'(\mb r,0)$. We use   a neural network model with two hidden layers, 3 exit nodes and 25 epochs. The training set contains realizations in the full phase diagram, except the transition region. The training set has $N_d=47$ realizations at each value of $\mu$ and $p$. With this model we predict $B$ for a validation set of $N_d=21$ new realizations for each value of $\mu$ and $p$.  (b) Counting of the predicted values of $B$ for $\mu=-t$ as a function of $p$.}\label{A2}
\end{figure}

On the other hand, in Fig.~\ref{A1}(b) we   consider the imaginary part of the pairing function $F''(\mb r,\omega=0)$. In this case, the ML model  trained with  clean systems   succeeds in  predicting  the topological phase diagram. This result indicates  that Majorana edge states display signatures in $F''(\mb r,\omega=0)$ in disordered  systems that are similar to the signatures in clean systems, as happens with the LDOS. Looking at the Bott index count for a fixed value of  $\mu$, however, we see that the model predicts $B=-1$ where this phase should not exist [see Fig.~\ref{A1}(c)].

In order to improve the accuracy of the phase diagrams, we train ML models in the whole phase diagram, except near  the transition from $B=2$ to $B=0$. Figure~\ref{A2} shows the result obtained for $F'(\mb r,\omega=0)$, which are now much more accurate. Similar accurate results were obtained when we trained ML models using $F''(\mb r,\omega=0)$ in the whole phase diagram.

\section{Conclusions \label{conclusion}}

We studied a model of a 2D superconductor in the presence of magnetic disorder and spin-orbit coupling. Depending on the magnetic dilution and the chemical potential, the system can be in  topologically trivial or nontrivial phases. We computed the Bott index $B$ for several disorder realizations and mapped out the topological phase diagram. There are phases with $B=0, -1, +2$, which correspond to the presence of zero, one, and two chiral Majorana  edge modes, respectively. Near the transition from  $B=+2$ to $B=0$, we find that several disorder realizations have an intermediate Bott index $B=+1$, which differ  from  $B=-1$  by the  chirality of the edge mode. 

Several experiments have reported signatures of Majorana edge states but cannot distinguish them from trivial edge modes. We showed that our model also leads to the presence of trivial edge modes. We then used machine learning algorithms to predict $B$ based on the local density of states at zero energy. We first trained an ML   model based on clean superconductors, for which the values of chemical potential at which the topological transitions take place are known analytically. The trained ML model predicted a topological phase diagram in fair agreement with the one obtained from the direct calculation of $B$. After that, we trained another ML model including disordered systems in the training set, excluding the region in parameter space around the transition. In this case, we considered a single node in the output layer of the neural network. Interestingly, the model was able to predict several realizations with $B=+1$ around the transition, even though the training set contained systems with $B=0,-1,+2$ only. The accuracy outside the transition region is higher than $99.5\%$. 

Importantly, the ML models correctly predict $B=0$ for trivial edge modes. 
Therefore, ML can be an important tool to distinguish between topologically trivial and nontrivial edge modes when the local density of states at zero energy is obtained from scanning tunneling spectroscopic experiments. This method can also distinguish between topological phases with different numbers of chiral edge states. We stress that the ML model trained with clean superconductors has a fair prediction of the topological phase diagram of disordered systems. Similarly,  ML algorithms trained with disordered theoretical models may be able to predict the topological invariant of a real superconducting sample, which is naturally more complex.   \new{ However, before applying the method to real world data, it would be important to generalize our approach and  train ML models with different geometries, other kinds of disorder, and model parameters that are close to the experimental setting.}

\begin{figure}[t]
\includegraphics[width=\columnwidth]{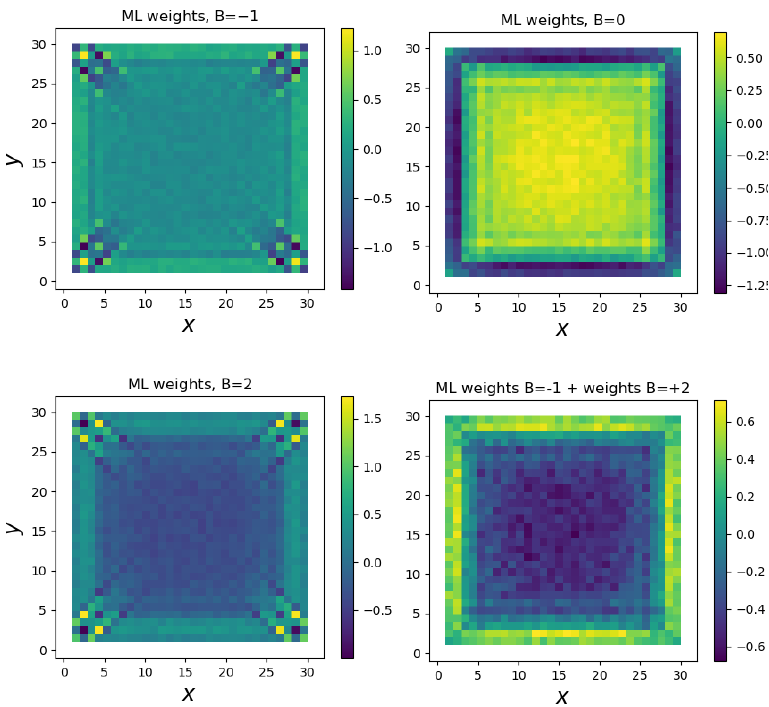}
\caption{ Weights $W$ in the linear function $Z=W X + b$ associated with the LDOS for every lattice site. We display the weights for the phases with $B=-1$ (top left), $B=0$ (top right), and $B=2$ (bottom left).  The sum of weights for $B=-1$ and $B=2$ is shown in the bottom right. }\label{F9}
\end{figure}

We also showed the presence of  unconventional odd-frequency pairing in the system. We trained another ML model with the local density of states at zero energy to predict the odd-frequency pairing at zero energy. We considered a simple linear regression model and obtained a fair prediction of the odd-frequency pairing. This result emphasizes the existence of a strong connection between the density of states and the imaginary part of the pairing function. It also demonstrates that ML can be used together with spectroscopic probes of superconductors to predict the strength of odd-frequency pairing.

\section{Acknowledgements}
We thank Eric Andrade  and Diego Ferreira for helpful discussions. This work was supported by the National Council for Scientific and Technological Development -- CNPq (F.N., A.C., R.C., R.G.P.) and by a grant from the Simons
Foundation (Grant No. 1023171, R.C., R.G.P.). Research at IIP-UFRN is supported by Brazilian ministries MEC and MCTI. 
We also acknowledge the High-Performance Computing Center (NPAD) at UFRN for providing computational resources. AC acknowledges  partial financial support by the Alagoas State Research Agency (FAPEAL) (Grant No. APQ$2022021000153$).

\appendix

\section{Weights in the ML model \label{apA}}

To gain insight into how  the ML model extracts features from the LDOS to classify the phases, here we discuss the weights in the neural network for the simple case of Fig. \ref{F4}, where we used no hidden layer. 
The exit layer has three nodes, which correspond to each of the classes $B=-1,0,+2$. Each node computes the linear function $Z=W X + b$, where $X$ is an array containing the values of zero-energy LDOS for all lattice sites, $W$ is an array with the weights of the neural network, and $b$ is a bias scalar. Both $W$ and $b$ are estimated during the training of the ML model. The value of $Z$ is then transformed to a probability through the softmax activation function. In Fig. \ref{F4}, the computed values of $b$ are $b=0.9625709, 2.04757, -2.566742$ for $B=-1,0,+2$, respectively. The weights $W$, computed for each $B$, are shown in   Fig. \ref{F9} for each position of the lattice. One can see that if the LDOS is higher in the bulk, the value of $Z$ for $B=0$ is also high, leading to a high probability that the sample has $B=0$. If, however, the LDOS is higher along the edges, the $Z$ values for $B=-1,+2$ increase. Interestingly, the weights  for $B=-1,+2$ display some
structure along the diagonals near the corners. The last plot shows the sum of $W$ for $B=-1$ with $W$ for $B=2$. In this case the diagonal structure disappears. This means that the structure
that the neural network  expects in the diagonals for $B=-1$ is complementary to the one for $B=+2$. We stress, however, that this analysis holds for a neural network without hidden layers. The presence of two
hidden layers can change the expected structure of the LDOS for each class. Nevertheless,  this analysis illustrates    how the edge states and structures near the corners can be used to predict the Bott index.

\bibliographystyle{apsrev4-1_control}
\bibliography{references}
\end{document}